\documentclass[twocolumn]{svjour3}

\usepackage{graphicx}
\usepackage{color}
\usepackage{newlfont}
\usepackage{amssymb}
\usepackage{amsmath}
\usepackage{bbm}
\usepackage{cancel}
\usepackage{latexsym}
\usepackage{bbm}
\usepackage{cite}
\usepackage{epstopdf}
\usepackage[utf8]{inputenc}
\usepackage{braket}

\newcommand{\beq}{\begin{eqnarray}}
\newcommand{\eeq}{\end{eqnarray}}
\newcommand{\be}{\begin{equation}}
\newcommand{\ee}{\end{equation}}
\newcommand{\sS}{\textrm{\tiny$S$}}
\newcommand{\sSl}{\textrm{\tiny$S'$}}
\newcommand{\sA}{\textrm{\tiny$A$}}
\newcommand{\lu}{\textrm{\tiny$\mathfrak{L\,U}$}}
\newcommand{\ul}{\textrm{\tiny$\mathfrak{U\,L}$}}
\newcommand{\LL}{\textrm{\tiny$\mathfrak{L}$}}
\newcommand{\UU}{\textrm{\tiny$\mathfrak{U}$}}
\newcommand{\sSig}{\textrm{\tiny $\Sigma$}}

\begin{document}
\title{Time Contraction Within Lightweight Reference Frames}
\author{Matheus F. Savi \and Renato. M. Angelo}
\institute{M. F. Savi and R. M. Angelo
\at Department of Physics, Federal University of Paraná, P.O.Box 19044, Curitiba, PR 81531-980, Brazil \\ \email{mfs11@fisica.ufpr.br, renato@fisica.ufpr.br}}

\maketitle

\begin{abstract}
The special theory of relativity teaches us that, although distinct inertial frames perceive the same dynamical laws, space and time intervals differ in value. We revisit the problem of time contraction using the paradigmatic model of a fast-moving laboratory within which a photon is emitted and posteriorly absorbed. In our model, however, the laboratory is composed of two independent parallel plates, each of which allowed to be sufficiently light so as to get kickbacks upon emission and absorption of light. We show that the lightness of the laboratory accentuates the time contraction. We also discuss how the photon frequency shifts upon reflection in a light moving mirror. Although often imperceptible, these effects will inevitably exist whenever realistic finite-mass bodies are involved. More fundamentally, they should necessarily permeate any eventual approach to the problem of relativistic quantum frames of reference.
\end{abstract}

%==========================================
\section{Introduction}\label{introduction}

Think of a universe with only a single system, say an electron. Is it possible to make physical assertions about its position, velocity, energy, mass, spin, charge, temperature, wave function? Some people would answer this question in the affirmative by arguing that one can always consider an immaterial reference system for which such assertions would make sense. In contrast, we defend that this position is not useful in physics because, within any scientific theory, we always look for a description that can be empirically verified by some observer. Now, the notion of ``observer'' does not need to be linked with a brain-endowed organism. It suffices to think of a laboratory equipped with apparatuses whose measurement outcomes make reference to a space-time structure, which is defined by rods and clocks rigidly attached to this laboratory. Being a physical entity, the laboratory itself is subjected to the laws of physics. This means that the very act of observing a system may imply a kickback to the laboratory, which then would become a non-inertial reference frame. This is the notion of {\em reference frame} we will keep in mind throughout this work: a coordinate system rigidly attached to a real laboratory which can physically interact with other systems. 

The concept of a physical reference frame is well known in classical mechanics. It appears in many text-books\cite{goldstein,marion}, for instance in the form of {\em the two-body problem}, where the dynamics of two interacting point masses is analysed while one of these particles is promoted to reference frame. As a result, Newton's second law for the relative physics is shown to depend on the two-body reduced mass, which incorporates the effects of the reference frame lightness. As far as the quantum domain is concerned, it has been shown that nonrelativistic quantum mechanics can be consistently formulated in terms of {\em genuine quantum reference frames} \cite{ak84,spekkens07,angelo11,angelo12}, i.e., interacting quantum particles playing the role of reference frames. More recently, one of us and a collaborator pointed out a framework, involving quantum reference frames, in which quantum theory is both manifestly covariant under Galilei boosts and compatible with Einstein's equivalence principle~\cite{angelo15}.

The question then arises as to whether relativistic quantum theory as well, as described by Dirac's equation, will preserve its covariance upon Lorentz transformations in a scenario where the moving frame itself is a {\em relativistic quantum particle} (i.e., a particle prepared in a quantum state with large mean momentum relative to a given inertial reference frame). While the answer to this query is expected to be given in the positive, it seems to us, in light of the efforts demanded in the work~\cite{angelo15}, that the demonstration of such a point is a rather complicated task. In fact, to the best of our knowledge, no related study has been reported so far and, to be honest, neither do we conceive by now a clear proposal on how to conduct such an investigation. 

We then propose to assess here, as a first prototype, a much simpler situation in which the relativistic quantum particles involved can be thought of as being prepared in minimal uncertainty wave packets with large mean momentum. As a further simplification, we may assume that the wave packets remain significantly localized during the experiment and that no particle-antiparticle excitation is activated throughout, so that we can apply the Ehrenfest theorem to the dynamics and thus effectively treat the quantum systems as classical particles. We nevertheless use the Lorentz transformations to look at the physics ``seen'' by one of these relativistic particles while the system evolves in time. All this makes the aforementioned complex problem involving relativistic quantum reference frames reduce to the problem of describing the physics from the perspective of a {\em finite-mass relativistic classical particle}, a model that has not been investigated so far either.

Being more specific, we will phrase our approach in terms of a convenient adaptation of the famous problem in which light travels within a fast-moving train \cite{halliday,tipler}. In the usual version of this problem, light is emitted from the floor of the train, reflects in the roof, and is absorbed at the same point in the floor. An observer within the train measures, using a single clock, a time interval $\Delta t_{\sSl}$ between the two events (emission and absorption in the floor), which occur at the same position in his inertial reference frame $S'$. An observer in an external inertial reference frame $S$ measures, using two synchronized clocks placed in different locations, a time interval $\Delta t_{\sS}$. According to the laws of the special relativity, these time intervals are related by the formula
\be\label{usualDtDt'}
\Delta t_{\sSl}= \Delta t_{\sS}/\gamma_u,
\ee
where $\gamma_u=\left(1-u^2/c^2\right)^{\text{\tiny $-\frac{1}{2}$}}$, $c$ is the speed of light in vacuum, and $u$ is the speed of the train relative to $S$. Given that $u<c$, then $\gamma_u>1$ and $\Delta t_{\sSl}<\Delta t_{\sS}$. This implies a contraction of the internal time interval relatively to the external one. (The time contraction\footnote{Usually, Eq.~\eqref{usualDtDt'} is expressed as $\Delta t_{\sS}=\gamma_u\,\Delta t_{\sSl}$ and referred to as a statement of {\em time dilation}. In this form, the focus is on the external time interval. In this work, we opt to focus the attention onto the physics of the internal reference frame, so that we use $\Delta t_{\sSl}=\Delta t_{\sS}/\gamma_u$ and refer to the phenomenon as {\em time contraction}.} is by now a well established fact, having been demonstrated in several experiments~\cite{kunding63,hafele72,bailey77,mcgowan93,saathoff03,reinhardt07,lammerzah07,najjari08,dube14,botermann14}.) Here we will consider a sort of ``microscopic elastic version'' of this problem in which the light beam is replaced with a single photon and the {\em rigid train} with two very light plates, which can move independently. In our model, the upper plate will be a mirror and the lower plate will play the role of moving reference frame. The motivation behind this scheme is to understand how the formula \eqref{usualDtDt'} changes in a regime in which the moving system is allowed to get kickbacks upon emission and absorption of light, as would do a quantum particle. Although, on the one hand, we may suspect that any eventual correction must be negligible, on the other, conservation laws ensure that it is fundamentally unavoidable.

In spite of all the idealizations of this model, specially the tacit use of Ehrenfest's theorem to approximate quantum systems by classical particles, thus avoiding the mathematical complications that would inevitably derive from the uncertainty principle, we still expect to get some insight on the sort of phenomenon we should meet from the perspective of relativistic quantum particles. After all, be localized as a classical particle or delocalized as a quantum wave, any finite-mass system is compulsorily submitted to kickbacks deriving from the conservation laws.

%==========================================
\section{Model}\label{model}

\begin{figure}[ht]
\centering
\includegraphics[scale=0.1]{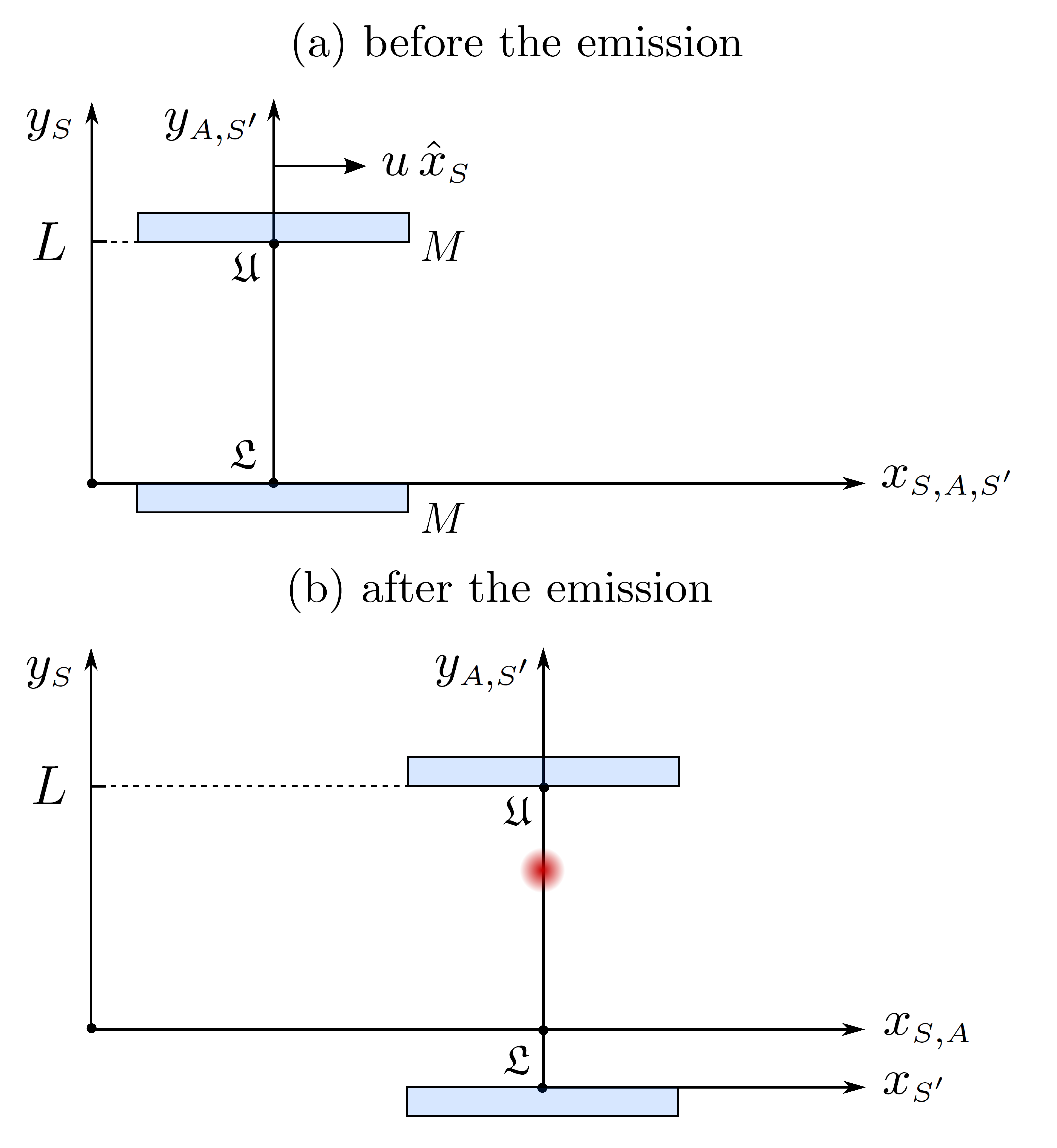}
\caption{\small (a) Two plates of mass $M$ move independently with velocity $u\,\hat{x}_{\sS}$ relative to an external inertial reference frame $S$. The upper plate is a mirror and the lower one, which is equipped with a source of photons and a cartesian system $[xy]_{\sSl}$, assumes the role of moving reference frame $S'$. An auxiliary reference frame $A$ moves with velocity $u\,\hat{x}_{\sS}$ relative to $S$. (b) After a photon (red spot) is emitted from the point $\mathfrak{L}$ at the lower plate towards the point $\mathfrak{U}$ at the upper mirrored plate, $S'$ starts to move along the $y_{\sS,\sA}$ axes.}
\label{fig1}
\end{figure}

Let us consider the framework illustrated in Fig. \ref{fig1}. Two parallel plates, each of mass $M$, move with velocities $\mathbf{V}_{\sS}^{\UU}=\mathbf{V}_{\sS}^{\LL}=u\, \hat{x}_{\sS}=(u,0)$ relative to an inertial reference frame $S$, where $\hat{x}_{\sS}$ is a unit vector associated with the cartesian coordinate system $[xy]_{\sS}$ that defines $S$. The superscripts $\mathfrak{U}$ and $\mathfrak{L}$ refer to points located at the {\em upper} and {\em lower} plates, respectively. These indexes are also used to name the plates themselves. Rigidly attached to the point $\mathfrak{L}$ of the lower plate is the origin of a cartesian system $[xy]_{\sSl}$, which then defines the {\em moving reference frame} $S'$. The upper plate is an ideal mirror. For future convenience, we also consider an {\em auxiliary} reference frame $A$, equipped with a cartesian system $[xy]_{\sA}$, that moves with constant velocity $u\,\hat{x}_{\sS}$ relative to $S$ and is perfectly aligned with $[xy]_{\sSl}$. Hence, the initial velocities of the lower plate and the mirror relative to $A$ are $\mathbf{V}_{\sA}^{\LL}=\mathbf{V}_{\sA}^{\UU}=(0,0)$. The velocity of the lower plate relative to its own coordinate system is $\mathbf{V}_{\sSl}^{\LL}=(0,0)$ and the velocity of the mirror relative to the lower plate is $\mathbf{V}_{\sSl}^{\UU}=(0,0)$. 

When a photon is emitted from the point $\mathfrak{L}$ and moves towards the point $\mathfrak{U}$, the lower plate gets a kickback and starts to move along the $y_{\sS,\sA}$ axes, as shown in Fig. \ref{fig1}(b). Notice that from the perspective of $S$ the motion of $S'$ is two-dimensional, whilst for $A$ it is one-dimensional. This is the reason why $A$ is useful. From now on, besides considering the velocities of the plates $\mathbf{V}_{\sSig}^{\LL,\UU}$ relative to a given reference frame $\Sigma$, with $\Sigma=S,S',A$, we also look at the photon velocity $\mathbf{v}_{\sSig}$ relative to $\Sigma$. 

At this moment, it is worth noticing that the kickback imposed on $S'$ by the photon emission makes $S'$ turn into a non-inertial reference frame only for an insignificant lapse of time. This is so because the photon is quite a peculiar entity that cannot be accelerated; either it does not exist (before the emission) or it exists and moves with speed $c$ (after the emission). As a consequence, we have to admit that the velocity of $S'$ relative to $S$ changes from a given constant vector to another constant vector {\em instantaneously}. It then follows that $S'$ is effectively inertial during all the relevant time intervals, so that we can safely apply the Lorentz transformations.

To determine the velocity acquired by $S'$ due to the emission of the photon, we apply the relativistic conservation laws from the perspective of the inertial reference frame $A$. Energy and momentum conservation laws imply, respectively, that
\begin{subequations}
\be
Mc^2=\frac{M_{\LL\sA}c^2}{\sqrt{1 - \mathrm{V}_{\LL\sA}^2/c^2}} + h \nu_{\sA}
\ee
and
\be
\frac{h \nu_{\sA}}{c} = \frac{M_{\LL\sA} \mathrm{V}_{\LL\sA}}{\sqrt{1 - \mathrm{V}_{\LL\sA}^2/c^2}},
\ee
\end{subequations}
where $M_{\LL\sA}$ $(M)$ is the rest mass of the lower plate after (before) the photon emission, $h$ is the Planck constant, $\nu_{\sA}$ is the photon frequency relative to $A$, and $-\mathrm{V}_{\LL\sA}\hat{y}_{\sA}$ is the velocity of the lower plate relative to $A$. Solving the above equations for $\mathrm{V}_{\LL\sA}$ and $M_{\LL\sA}$ yields
\begin{subequations}
\be
\frac{\mathrm{V}_{\LL\sA}}{c} = \frac{\epsilon}{1 - \epsilon} \qquad \textrm{and} \qquad \frac{M_{\LL\sA}}{M}=\sqrt{1 - 2\epsilon},
\label{ml,vl}
\ee
with
\be\label{epsilon}
\epsilon \equiv \frac{h \nu_{\sA}}{Mc^2}.
\ee
\end{subequations}
Because the photon always carries a non-zero energy and $\mathrm{V}_{\LL\sA}<c$, the parameter $\epsilon$ has to be bounded as $0 < \epsilon < 1/2$. In ordinary instances involving low energy photons and heavy plates, one has that $\epsilon\ll 1/2$. In this regime, it follows that $Mc^2 \simeq M_{\LL\sA}c^2+h\nu_{\sA}$, which is an expression of the mass-energy conservation expected for decay processes in nonrelativistic regime \cite{angelo15,greenberger01}. Throughout this paper, however, we keep $\epsilon$ arbitrary in the range $\left(0,\frac{1}{2}\right)$. 

In order to link the results of distinct inertial reference frames, we use the Lorentz transformations. Let $\Sigma'$ be a reference frame moving with constant velocity $v\,\hat{x}_{\sSig}$ relative to $\Sigma$, which is an inertial reference frame. In this instance, the Lorentz transformations can be written as
\be\label{LT}
\left(\begin{matrix}\Delta x_{\sSig'} \\ \\ \Delta y_{\sSig'} \\ \\ \Delta t_{\sSig'}  \end{matrix}\right)=\left( \begin{matrix} \gamma_{v} & & 0 & & -v\gamma_v \\ \\ 0 & & 1 & & 0 \\ \\ -\frac{v\gamma_v}{c^2} && 0 && \gamma_v\end{matrix} \right)\left(\begin{matrix}\Delta x_{\sSig} \\ \\ \Delta y_{\sSig} \\ \\ \Delta t_{\sSig}  \end{matrix}\right),
\ee
where $\Delta r=r_{\textrm{\tiny event (ii)}}-r_{\textrm{\tiny event (i)}}$ is an interval that measures the ``distance'' between events (ii) and (i) in terms of the variable $r$, with $r=x,y,t$. Thus, $r_{\textrm{\tiny event (i)}}$ denotes the ``location'' at which the event (i) occurred in the $r$-space, and similarly for $r_{\textrm{\tiny event (ii)}}$. When the motion of $\Sigma'$ occurs along a different axis, Eqs. \eqref{LT} has to be accordingly adapted. For the inverse transformations, one has that
\be\label{ILT}
\left(\begin{matrix}\Delta x_{\sSig} \\ \\ \Delta y_{\sSig} \\ \\ \Delta t_{\sSig}  \end{matrix}\right)=\left( \begin{matrix} \gamma_{v} & & 0 & & v\gamma_v \\ \\ 0 & & 1 & & 0 \\ \\ \frac{v\gamma_v}{c^2} && 0 && \gamma_v\end{matrix} \right)\left(\begin{matrix}\Delta x_{\sSig'} \\ \\ \Delta y_{\sSig'} \\ \\ \Delta t_{\sSig'}  \end{matrix}\right),
\ee
We are now ready to compute the time elapsed since the emission of the photon at the point $\mathfrak{L}$ until its absorption at the same point. For convenience, we divide the kinematics in two parts. 

%==========================================
\subsection{Photon's rise $(\mathfrak{L}\to\mathfrak{U})$}\label{rise}

We first calculate the time interval $\Delta t_{\sA}^{\lu}$ referring to the photon's rise from $\mathfrak{L}$ to $\mathfrak{U}$ from the perspective of $A$. In this case, the events to be considered are (i) the photon emission at the point $\mathfrak{L}$ located in the lower plate and (ii) the photon absorption at $\mathfrak{U}$, which is a point located in the upper mirrored plate (see Fig. \ref{fig1}). From the discussion above, one has that the velocity of the lower plate after the photon emission is $\mathbf{V}_{\sA}^{\LL}=\left(0,-\mathrm{V}_{\LL\sA}\right)$, whereas the velocity of the mirror is $\mathbf{V}_{\sA}^{\UU}=\left(0,0\right)$. Since the photon speed is the same in all reference frames, we have that $\mathbf{v}_{\sA}=(0,c)$. 

Concerning space-time intervals, for the events in question it is clear that $\Delta x_{\sA}^{\lu}=0$, $\Delta y_{\sA}^{\lu}=L$, and $\Delta t_{\sA}^{\lu}=L/c$. Then, we can apply the transformations \eqref{LT} and \eqref{ILT}, with pertinent adaptations, to obtain the frame conversions $A\to S'$ and $S'\to S$. The results can be expressed as
\beq
\begin{array}{lll}
\Delta x_{\sSl}^{\lu} = 0,& \qquad & \displaystyle\Delta x_{\sS}^{\lu} = \frac{\gamma_u\, u\, \Delta t_{\sSl}^{\lu}}{(1 - 2\epsilon)^{\text{\tiny $-\frac{1}{2}$}}}, \\ \\
\displaystyle\Delta y_{\sSl}^{\lu} = \frac{\Delta y_{\sA}^{\lu}}{(1 - 2\epsilon)^{\text{\tiny $\frac{1}{2}$}}}, 
& \qquad & \displaystyle\Delta y_{\sS}^{\lu} = \frac{\Delta y_{\sSl}^{\lu}}{(1 - 2\epsilon)^{\text{\tiny $-\frac{1}{2}$}}}, \\ \\
\displaystyle\Delta t_{\sSl}^{\lu} = \frac{\Delta t_{\sA}^{\lu}}{(1 - 2\epsilon)^{\text{\tiny $\frac{1}{2}$}}},& \qquad & \displaystyle\Delta t_{\sS}^{\lu} = \frac{\gamma_u\, \Delta t_{\sSl}^{\lu} }{(1 - 2\epsilon)^{\text{\tiny $-\frac{1}{2}$}}}. 
\end{array}\label{Dr_fr}
\eeq
The last relation above shows that the usual dilation factor $\gamma_u$ is reduced by a {\em recoil factor} $\sqrt{1-2\epsilon}$. In particular, no dilation will occur when $h\nu_{\sA}=Mu^2/2$, since in this case $\gamma_u \sqrt{1 - 2\epsilon}=1$. Of course, this regime cannot be reached when low energy photons and heavy plates are involved. 

%==========================================
\subsection{Photon's descent $(\mathfrak{U}\to \mathfrak{L})$}\label{descent}

Before considering the descent of the photon from $\mathfrak{U}$ to $\mathfrak{L}$, we need to look at the scattering of the photon by the mirror. From $A$'s perspective, the energy-momentum conservation in the {\em absorption process} at $\mathfrak{U}$, demands that
\begin{subequations}
\be
h\nu_{\sA}+Mc^2=\frac{M_{\UU\sA}c^2}{\sqrt{1 - \mathrm{V}_{\UU\sA}^2/c^2}}
\ee
and 
\be 
\frac{h \nu_{\sA}}{c} = \frac{M_{\UU\sA} \mathrm{V}_{\UU\sA}}{\sqrt{1 - \mathrm{V}_{\UU\sA}^2/c^2}},
\ee
\end{subequations}
whose solution is given by
\be
\frac{\mathrm{V}_{\UU\sA}}{c} = \frac{\epsilon}{1+\epsilon} \qquad \textrm{and} \qquad \frac{M_{\UU\sA}}{M}=\sqrt{1+2\epsilon},
\label{mu,vu}
\ee
with $\epsilon$ defined by Eq. \eqref{epsilon}. Here $\mathrm{V}_{\UU\sA}\hat{y}_{\sA}$ and $M_{\UU\sA}$ denote, respectively, the velocity and the rest mass of the mirror after the absorption of the photon. From a quantum mechanical viewpoint, the photon is absorbed by one atom of the mirror and is posteriorly emitted with a different frequency (as we will discuss latter). The time elapsed between these two events---the {\em lifetime} of the corresponding electronic transition---is denoted here by $\tau_{\sA}$.

During a time interval that comprises the photon rise and the atomic lifetime, the lower plate moves downwards a distance $\mathrm{V}_{\LL\sA}(\Delta t_{\sA}^{\lu}+\tau_{\sA})$ in $A$. By its turn, the mirror moves upwards a distance $\mathrm{V}_{\UU\sA}\tau_{\sA}$. From this moment on, the two events to be considered are (i) photon emission at $\mathfrak{U}$ and (ii) photon absorption at $\mathfrak{L}$. The time equations for the photon and the lower plate can be respectively written as $y_{\sA}(t_{\sA})=(L+\mathrm{V}_{\UU\sA}\tau_{\sA})-ct_{\sA}$ and $Y_{\sA}^{\LL}(t_{\sA})=-\mathrm{V}_{\LL\sA}(\Delta t_{\sA}^{\lu}+\tau_{\sA}+t_{\sA})$, where $t_{\sA}$ is the time elapsed since the emission of the photon at $\mathfrak{U}$. Equating these expressions, we can easily determine the time elapsed between the two events:
\be
\Delta t_{\sA}^{\ul}=\frac{L/c}{(1-2\epsilon)}+\frac{2\epsilon\,\tau_{\sA}}{(1+\epsilon)(1-2\epsilon)}.
\label{DtA_ul}
\ee
Now, using the Lorentz transformations and the above results, we can compute the total time elapsed from the emission at $\mathfrak{L}$ until the absorption at $\mathfrak{L}$ in all reference frames:
\beq
\Delta t_{\sA}&=&\Delta t_{\sA}^{\lu}+\tau_{\sA}+\Delta t_{\sA}^{\ul}=\frac{2L}{c}f(\epsilon)+\tau_{\sA}\,g(\epsilon), \nonumber \\ 
\Delta t_{\sS}&=&\gamma_u\,\Delta t_{\sA}, \\ 
\Delta t_{\sSl}&=&\gamma_{\textrm{\tiny $(-\mathrm{V}_{\LL\sA})$}}\left[\Delta t_{\sA}-\frac{(-\mathrm{V}_{\LL\sA})\Delta y_{\sA}}{c^2}\right]=\Delta t_{\sA}/\gamma_{\textrm{\tiny $\mathrm{V}_{\LL\sA}$}},\nonumber
\eeq
where we have used $\Delta y_A=-\mathrm{V}_{\LL\sA}\Delta t_{\sA}$ to derive the last equality and introduced the functions $f(\epsilon)\equiv\frac{1-\epsilon}{1-2\epsilon}$ and $g(\epsilon)\equiv\frac{1+\epsilon-2\epsilon^2}{1-\epsilon-2\epsilon^2}$ for the sake of notational compactness. To conclude, we can write
\be
\Delta t_{\sSl}=\frac{\Delta t_{\sS}}{\gamma_u \gamma_{\textrm{\tiny $\mathrm{V}_{\LL\sA}$}}}=\left(\frac{\sqrt{1-2\epsilon}}{1-\epsilon}\right)\Delta t_{\sS}/\gamma_u.
\label{newDtDt'}
\ee
It is interesting to notice that this relation does not depend on any information regarding the photon scattering in the mirror (as for instance $\tau_{\sA}$); such information is encoded only on $\Delta t_{\sA}$. In addition, the first equality above gives an intuitive relation: $\Delta t_{\sSl}$ connects with $\Delta t_{\sS}$ through Lorentz factors referring to both the horizontal motion and the recoil of $S'$ relative to $S$.

So far, our results have been expressed in terms of $\nu_{\sA}$, which is the frequency observed from the auxiliary reference frame $A$ during the photon rise. Now we want to abandon $A$ and rewrite our results in terms of $\nu_{\sSl}$, which is the frequency measured in $S'$. To this end, we apply the {\em longitudinal relativistic Doppler effect}~\cite{halliday,tipler}. It is clear that, when the photon is rising, ``the source'' at $\mathfrak{L}$ separates with speed $\mathrm{V}_{\LL\sA}$ from ``the detector'', which is fixed say at the origin $y_{\sA}=0$ of $A$. It follows that the frequency values during the photon rise as seen by $A$ and $S'$ are related as
\be
\nu_{\sA}=\nu_{\sSl}\sqrt{\frac{1-\mathrm{V}_{\LL\sA}/c}{1+\mathrm{V}_{\LL\sA}/c}}=\nu_{\sSl}\sqrt{1-2\epsilon}.
\label{nuA}
\ee
This is still not the solution to the problem because the r.h.s. term in the last equality depends on $\nu_{\sA}$ through the relation $\epsilon=\frac{h\nu_{\sA}}{Mc^2}$ defined in Eq. \eqref{epsilon}. Using this relation, we can solve Eq. \eqref{nuA} for $\nu_{\sA}$ so as to obtain
\be
\nu_{\sA}=\nu_{\sSl}\left(\sqrt{1+\varepsilon^2}-\varepsilon \right), \qquad \varepsilon\equiv \frac{h\nu_{\sSl}}{Mc^2},
\label{nuAE}
\ee
where $\varepsilon$ has substituted $\epsilon$ in the role of ``significant dimensionless parameter.'' 

We are now in position to finish our calculations. By direct inspection of Eqs. \eqref{nuA} and \eqref{nuAE}, we learn how to express $\sqrt{1-2\epsilon}$ as a function of $\varepsilon$. With that, we come back to Eq. \eqref{newDtDt'} to derive, after some algebraic manipulations, our final result:
\be
\Delta t_{\sSl}=\frac{\Delta t_{\sS}/\gamma_u}{\sqrt{1+\varepsilon^2}}.
\label{final}
\ee
In comparing this result with Eq. \eqref{usualDtDt'}, which is obtained in the usual context of a rigid infinite-mass laboratory, one can readily regard $\sqrt{1+\varepsilon^2}$ as a correction factor deriving from considering a nonrigid finite-mass laboratory. Indeed, Eq. \eqref{final} reduces to Eq. \eqref{usualDtDt'} whenever $Mc^2\gg h\nu_{\sSl}$. It is worth noticing that $\varepsilon$ will be small even in extreme scenarios, as for example when the laboratory is thought of as being formed by only two hydrogen atoms (e.g., a H$_2$ molecule), one representing the lower plate and the other the mirror. Taking $M\cong 1.0$ u for the mass of each plate gives $Mc^2\simeq 930$ MeV. If we consider the highest-energy photon that an hydrogen atom could eventually emit, we can estimate that $h\nu_{\sSl}\simeq 14$ eV. It follows that $\varepsilon \simeq 1.5 \times 10^{-8}$, which makes $\varepsilon^2$ negligible in Eq. \eqref{final}.

Before concluding, a quick remark is opportune with respect to the phenomenon of frequency change upon reflection in a {\em light} moving mirror. Consider a mirror plate of mass $M$ moving with velocity $v\hat{y}_{\sS}$ with respect to an inertial reference frame $S$. The plane of the mirror is always perpendicular to $\hat{y}_{\sS}$. A photon with velocity $c\hat{y}_{\sS}$ and frequency $\nu_i$ impinges on the mirror and reflects with velocity $-c\hat{y}_{\sS}$ and frequency $\nu_r$. After the photon reflection, the mirror moves with speed $v'$. The conservation laws for this scenario,
\begin{subequations}
\be
h\nu_i+\frac{Mc^2}{\sqrt{1-\frac{v^2}{c^2}}}=\frac{Mc^2}{\sqrt{1-\frac{v'^2}{c^2}}}+h\nu_r
\ee
and
\be
\frac{h\nu_i}{c}+\frac{Mv}{\sqrt{1-\frac{v^2}{c^2}}}=\frac{Mv'}{\sqrt{1-\frac{v'^2}{c^2}}}-\frac{h\nu_r}{c},
\ee
\end{subequations}
require that the frequency change $\nu_i\to\nu_r$ upon reflection be described, to the $S$ perspective, as
\begin{subequations}
\be
\nu_r=\nu_i\left(\frac{1-\beta}{1+\beta}\right)\Gamma,
\ee
where 
\be
\Gamma \equiv \left(1+2\epsilon_i \sqrt{\frac{1-\beta}{1+\beta}}\right)^{-1},
\ee
\end{subequations}
$\beta=v/c$ and $\epsilon_i\equiv\frac{h\nu_i}{Mc^2}$. For infinite-mass mirrors, $\epsilon_i\to 0$ and $\Gamma\to 1$, in which case we recover the usual formula for light reflection in a moving mirror~\cite{gjurchinovski04}. Notice that even if $v=0$, a correction $\Gamma=(1+2\epsilon_i)^{-1}$ will be present due to the lightness of the mirror. Of course, this correction could be implemented in the problem under scrutiny in this work, if required. In our approach, however, this was not necessary because the results have been exhibited in terms of the initial frequency of the photon, the one defined in the very first emission at the lower plate. 

%==========================================
\section{Conclusion}\label{conclusion}

Once we admit that a laboratory is never rigorously isolated from the observed system, with which it physically interacts, then we may wonder whether such a reference frame can indeed be regarded as inertial. Usually, one can maintain this position only to some degree of approximation. To obtain a more reliable description in such scenarios, and posteriorly assess the quality of those approximations, we often appeal to an {\em absolute} inertial reference frame from which we describe the physics of both the system and the laboratory. With this strategy, we then derive the pseudo physical laws, the ones that govern the behaviour of the system within the non-inertial laboratory. In the last decade, some authors reported work on Galilei boosts in quantum systems, thus discussing the physics seen from the perspective of quantum reference frames. To our perception, however, fundamental studies involving lightweight reference frames still need to be done in the fields of special relativity and relativistic quantum mechanics. 

The present work aims at contributing to this discussion by re-examining a well established physical phenomenon as seen by a relativistic lightweight particle. We ask how the usual formula governing the time contraction changes when light is emitted and absorbed within a finite-mass laboratory. In our model, we replace the traditional rigid massive train with two fast-moving parallel light plates, the lower of them playing the role of a reference frame $S'$ moving relatively to an inertial reference frame $S$. A single photon with energy $h\nu_{\sSl}$ is emitted from the lower plate, gets reflected in the upper plate (a mirror), and is finally absorbed at the emission point in the lower plate. Because of the relativistic energy-momentum conservation law, upon emission or absorption of the photon, the lower plate gets a kickback and instantaneously changes its velocity relative to $S$. Direct application of the Lorentz transformations allows us to predict, as our main result, that the contraction of the internal time $\Delta t_{\sSl}$ in relation to the external one, $\Delta t_{\sS}$, will be {\em accentuated} by the recoil of the plate. Even though the correction will often be irrelevant in practice, it will certainly be present whenever finite-mass laboratories are involved. (Hopefully, this effect will be accessible to future technologies!) Our result gives an interesting example of how a well-established effect of special relativity manifests itself in a scenario involving tiny reference frames. This anticipates the sort of phenomenon that we may find within quantum reference frames moving with relativistic speeds. 

\begin{center}
{\bf Acknowledgements}
\end{center}
We gratefully acknowledge A. D. Ribeiro, G. M. Kremer, and C. A. Duarte for discussions. M.F.S. and R.M.A. acknowledge financial support from CAPES and the National Institute for Science and Technology of Quantum Information (INCT-IQ, CNPq/Brazil), respectively. The final publication is available at Springer via\\ http://dx.doi.org/10.1007/s13538-017-0501-4

%\end{document}

%==========================
%\section*{References}


\begin{thebibliography}{99}
\bibitem{goldstein} H. Goldstein, C. Poole, J. Safko, Classical Mechanics, 3rd edn, (Addison-Wesley, New York, 2001)
\bibitem{marion} J.B. Marion, S.T. Thornton, Classical Dynamics of Particles and Systems, 5th edn, (Brooks/Cole -- Thomson Learning, Belmont, 2003)
\bibitem{ak84} Aharonov, Y., Kaufherr T.: Quantum frames of reference. Phys. Rev. D. {\bf 30}, 368 (1984)
\bibitem{spekkens07} Bartlett, S.D., Rudolph, T., Spekkens, R.W.: Reference frames, superselection rules, and quantum information. Rev. Mod. Phys. {\bf 79}, 555 (2007)
\bibitem{angelo11} Angelo, R.M., Brunner, N., Popescu, S., Short, A.J., Skrzypczyk, P.: Physics within a quantum reference frame. J. Phys. A: Math. Theor. {\bf 44}, 145304 (2011)
\bibitem{angelo12} Angelo, R.M., Ribeiro, A.D.: Kinematics and dynamics in noninertial quantum frames of reference. J. Phys. A: Math. Theor. {\bf 45}, 465306 (2012)
\bibitem{angelo15} Pereira, S.T., Angelo, R.M.: Galilei covariance and Einstein's equivalence principle in quantum reference frames. Phys. Rev. A {\bf 91}, 022107 (2015)
\bibitem{halliday} D. Halliday, R. Resnick, Fundamentals of Physics, 3rd edn, (John Wiley \& Sons, New York, 1998)
\bibitem{tipler} P.A. Tipler, G. Mosca, Physics for Scientists and Engineers, vol. 3, 6th edn, (W. H. Freeman, New York 2007)
\bibitem{kunding63} K\"unding, W.: Measurement of the Transverse Doppler Effect in an Accelerated System. Phys. Rev. {\bf 129}, 2371 (1963)
\bibitem{hafele72} Hafele, J.C., Keating, R.E.: Around-the-World Atomic Clocks: Predicted Relativistic Time Gains. Science {\bf 177} (4044), 168 (1972)
\bibitem{bailey77} Bailey, J., Borer, K., Combley, F., Drumm, H., Krienen, F., Lange, F., Picasso, E., von Ruden, W., Farley, F.J.M., Field, J.H., Flegel, W., Hattersley, P.M.: Measurements of relativistic time dilatation for positive and negative muons in a circular orbit. Nature {\bf 268}, 301 (1977)
\bibitem{mcgowan93} McGowan, R.W., Giltner, D.M., Sternberg, S.J., Lee, S.A.: New measurement of the relativistic Doppler shift in neon. Phys. Rev. Lett. {\bf 70}, 251 (1993)
\bibitem{saathoff03} Saathoff, G., Karpuk, S., Eisenbarth, U., Huber, G., Krohn, S., Horta, R.M., Reinhardt, S., Schwalm, D., Wolf, A., Gwinner, G.: Improved Test of Time Dilation in Special Relativity. Phys. Rev. Lett. {\bf 91}, 190403 (2003)
\bibitem{reinhardt07} Reinhardt, S., Saathoff, G., Buhr, H., Carlson, L.A., Wolf, A., Schwalm, D., Karpuk, S., Novotny, C., Huber, G., Zimmermann, M., Holzwarth, R., Udem, T., H\"ansch, T.W., Gwinner, G.: Test of relativistic time dilation with fast optical atomic clocks at different velocities. Nature Phys. {\bf 3}, 861 (2007)
\bibitem{lammerzah07} L\"ammerzah C: Special relativity: A matter of time. Nature Phys. {\bf 3}, 831 (2007)
\bibitem{najjari08} Najjari, B., Surzhykov, A., Voitkiv, A.B.: Relativistic time dilation and the spectrum of electrons emitted by 33-TeV lead ions penetrating thin foils. Phys. Rev. A {\bf 77}, 042714 (2008)
\bibitem{dube14} Dub\'e, P., Madej, A.A., Tibbo, M., Bernard, J.E.: High-Accuracy Measurement of the Differential Scalar Polarizability of a $^{88}\mathrm{Sr}^+$ Clock Using the Time-Dilation Effect. Phys. Rev. Lett. {\bf 112}, 173002 (2014)
\bibitem{botermann14} Botermann, B., Bing, D., Geppert, C., Gwinner, G., H\"ansch, T.W., Huber, G., Karpuk, S., Krieger, A., K\"uhl, T., N\"ortersh\"auser, W., Novotny, C., Reinhardt, S., S\'anchez, R., Schwalm, D., St\"ohlker, T., Wolf, A., Saathoff, G.: Test of Time Dilation Using Stored $\mathrm{Li}^+$ Ions as Clocks at Relativistic SpeedPhys. Rev. Lett. {\bf 113}, 120405 (2014)
\bibitem{greenberger01} Greenberger, D.M.: Inadequacy of the Usual Galilean Transformation in Quantum Mechanics. Phys. Rev. Lett. {\bf 87}, 100405 (2011)
\bibitem{gjurchinovski04} Gjurchinovski, A.: Reflection of light from a uniformly moving mirror. Am. J. Phys. {\bf 72}, 1316 (2004)
\end{thebibliography}
\end{document}